%% file: main.tex
\documentclass[10pt,a4paper,twocolumn,superscriptaddress,aps,prd,longbibliography,nofootinbib]{revtex4-2}
\usepackage[modulo]{lineno}
\usepackage{amsmath}
\usepackage{mathtools}
\usepackage{amsfonts}
\usepackage{amssymb}
\usepackage{bbold}
\usepackage{graphicx}
\usepackage[usenames,dvipsnames]{color}
\usepackage{float}
\usepackage{multirow}
\usepackage{soul}
\usepackage[normalem]{ulem}

\usepackage[nohyperlinks, printonlyused, nolist]{acronym}

\usepackage[range-units=single,separate-uncertainty]{siunitx}
\usepackage{verbatim} 

\graphicspath{ {./images/} }

\usepackage{blindtext}

\begin{document}

\author{Jonas Erhardt}
\affiliation{Physikalisches Institut, Universit\"at W\"urzburg, D-97074 W\"urzburg, Germany}
\affiliation{W\"urzburg-Dresden Cluster of Excellence ctd.qmat, Universit\"at W\"urzburg, D-97074 W\"urzburg, Germany}

\author{Sven Schemmelmann}
\affiliation{Physikalisches Institut, Universit\"at M\"unster, Wilhelm–Klemm–Straße 10, 48149 M\"unster, Germany}

\author{Fabian Sch\"ottke}
\affiliation{Physikalisches Institut, Universit\"at M\"unster, Wilhelm–Klemm–Straße 10, 48149 M\"unster, Germany}

\author{Cedric Schmitt}
\affiliation{Physikalisches Institut, Universit\"at W\"urzburg, D-97074 W\"urzburg, Germany}
\affiliation{W\"urzburg-Dresden Cluster of Excellence ctd.qmat, Universit\"at W\"urzburg, D-97074 W\"urzburg, Germany} 

\author{Lukas Gehrig}
\affiliation{Physikalisches Institut, Universit\"at W\"urzburg, D-97074 W\"urzburg, Germany}
\affiliation{W\"urzburg-Dresden Cluster of Excellence ctd.qmat, Universit\"at W\"urzburg, D-97074 W\"urzburg, Germany}

\author{Bing Liu}
\affiliation{Physikalisches Institut, Universit\"at W\"urzburg, D-97074 W\"urzburg, Germany}
\affiliation{W\"urzburg-Dresden Cluster of Excellence ctd.qmat, Universit\"at W\"urzburg, D-97074 W\"urzburg, Germany} 

\author{Domenico Di Sante}
\affiliation{Department of Physics and Astronomy, University of Bologna, Bologna, Italy}

\author{Giorgio Sangiovanni}
\affiliation{W\"urzburg-Dresden Cluster of Excellence ctd.qmat, Universit\"at W\"urzburg, D-97074 W\"urzburg, Germany}
\affiliation{Institut f\"ur Theoretische Physik und Astrophysik, Universit\"at W\"urzburg, D-97074 W\"urzburg, Germany}

\author{J\"org Sch\"afer}
\affiliation{Physikalisches Institut, Universit\"at W\"urzburg, D-97074 W\"urzburg, Germany}
\affiliation{W\"urzburg-Dresden Cluster of Excellence ctd.qmat, Universit\"at W\"urzburg, D-97074 W\"urzburg, Germany}

\author{Peter Kr\"uger}
\affiliation{Institut für Festkörpertheorie, Universit\"at M\"unster, Wilhelm–Klemm–Straße 10, 48149 M\"unster, Germany}

\author{Markus Donath}
\affiliation{Physikalisches Institut, Universit\"at M\"unster, Wilhelm–Klemm–Straße 10, 48149 M\"unster, Germany}

\author{Ralph Claessen}
\email{e-mail: ralph.claessen@uni-wuerzburg.de}
\affiliation{Physikalisches Institut, Universit\"at W\"urzburg, D-97074 W\"urzburg, Germany}
\affiliation{W\"urzburg-Dresden Cluster of Excellence ctd.qmat, Universit\"at W\"urzburg, D-97074 W\"urzburg, Germany}

\date{\today}



\begin{abstract}
Orbital angular momentum (OAM) characterization is central to understanding quantum materials such as valley Hall and Rashba systems and, through its connection to Berry curvature, provides key insight into band topology. Conventional OAM detection via photoemission dichroism, however, is susceptible to interference artifacts and is largely restricted to occupied states.
Here, we demonstrate a real-space approach to map the $L_z$ OAM character using scanning tunneling microscopy in triangular atomic monolayers. Unlike photoemission, this method exploits interference effects arising from the $L_z$ orbital phase and the Bloch phase associated with neighboring atoms. This shifts the local density of states (LDOS) maxima to distinct Wyckoff positions between atoms, encoding $L_z$ information into characteristic real-space LDOS patterns.
We exemplify this approach in 2D quantum materials with contrasting $L_z$ sequences, namely Tl/Si(111) and In/SiC(0001) monolayers, \textcolor{black}{identifying $L_z$-dependent LDOS accumulation associated with distinct topological phases in both systems.}
Our results establish $L_z$-dependent charge localization as a proxy to atomic obstruction for energetically isolated states in triangular lattices, including transition-metal dichalcogenides.
\end{abstract}


\title{\textcolor{black}{Probing $L_z$ Orbital Angular Momentum via Wavefunction-Interference-Induced Charge Order}}
\maketitle

\input{Textbody}

\bigskip
\textbf{Acknowledgements} 
We are grateful for funding support from the Deutsche Forschungsgemeinschaft (DFG, German Research Foundation) under Germany's Excellence Strategy through the W\"urzburg-Dresden Cluster of Excellence on Complexity, Topology and Dynamics in Quantum
Matter ctd.qmat (EXC 2147, Project ID 390858490) as well as through the Collaborative Research Center SFB 1170 ToCoTronics (Project ID 258499086).
D.D.S acknowledges MUR funding within the FIS2 (n. 1236, 01-08-2023) Project no. FIS-2023-00144 (CUP J53C25001880001).
%

\bigskip
\textbf{Data Availability} 
The raw data generated in this study have been deposited in the WueData database \cite{Wuedata} under accession code (TBD).


{\noindent
	\textbf{Author contributions}
J.E. and S.S. carried out the sample growth. J.E. performed the photoelectron spectroscopy and scanning tunneling microscopy experiments, as well as their analysis. P.K. conducted the calculations for Tl/Si(111), while G.S. and D.D.-S. developed the theoretical framework based on phase interference. Experimental contributions were provided by F.S., C.S., L.G., B.L., J.S., M.D., and R.C., while G.S., D.D.-S. and P.K. contributed to the theoretical aspects. R.C. and M.D. supervised the project. J.E. wrote the manuscript with input from all other authors.
}

{\noindent
	\textbf{Competing interests}
	The authors declare no competing interests.
}



%

\end{document}

%% file: Textbody.tex
The topology of electronic band structures fundamentally shapes the macroscopic properties of crystalline solids, giving rise to phenomena such as the quantum spin Hall (QSH) effect, large anomalous Hall effect, and topological superconductivity \cite{Koenig2007,Hasan2010,Qi2011,Nayak2016,Reis2017}. These effects are rooted in the Berry curvature, the imaginary part of the quantum geometric tensor, which captures the topological winding of wavefunctions in momentum space and has therefore motivated the development of Berry-curvature-sensitive probes. A key conceptual advance was the establishment of a direct link between Berry curvature and orbital angular momentum (OAM) within the modern theory of orbital magnetization \cite{Schuler2020,Xiao2010}. This connection enables experimental access to topological properties via OAM-sensitive techniques, most prominently circular dichroism in angle-resolved photoemission spectroscopy (ARPES) \cite{Cho2018,Unzelmann2020,Beaulieu2020,Uenzelmann2021,Figgemeier2025,Erhardt2024}. However, dichroism-based ARPES remains intrinsically limited to occupied states and is often affected by final-state effects, geometric artifacts, and spatial averaging in multi-domain samples \cite{Erhardt2024,Moser2023,Kern2023}.
\begin{figure}[t!]
\includegraphics[width=1\columnwidth, height=25cm, keepaspectratio]{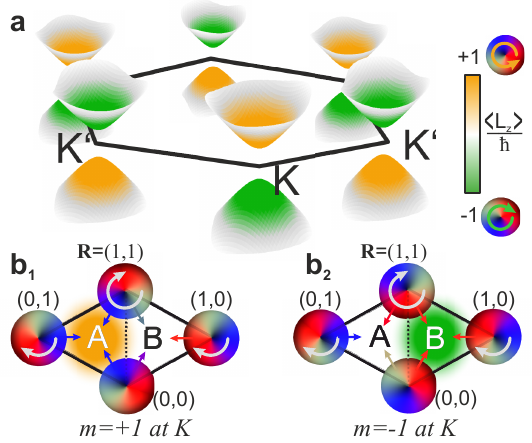}
\caption{\textbf{Real-space charge localization of $L_z$ eigenstates.} 
\textbf{a} Schematic band structure at the K/K' points of a triangular $p$-orbital lattice, yielding $L_z$ eigenstates due to an inversion symmetry breaking potential.
\textbf{b} Interference mechanism for $p_{\pm}$-orbitals on a triangular lattice: The site \textbf{R}-dependent Bloch phase (length of gray arrows) adds an offset to the orbital phase $m\phi(\textbf{R})$ (indicated by color sequence red-green-blue).
The net effect is either constructive or destructive interference for the $L_z$ eigenstates at A or B ($m=+1$ at K, equivalent to $m=-1$ at K').
}
\label{Fig1}
\end{figure}

\begin{figure*}[t!]
\includegraphics[width=18cm, height=25cm, keepaspectratio]{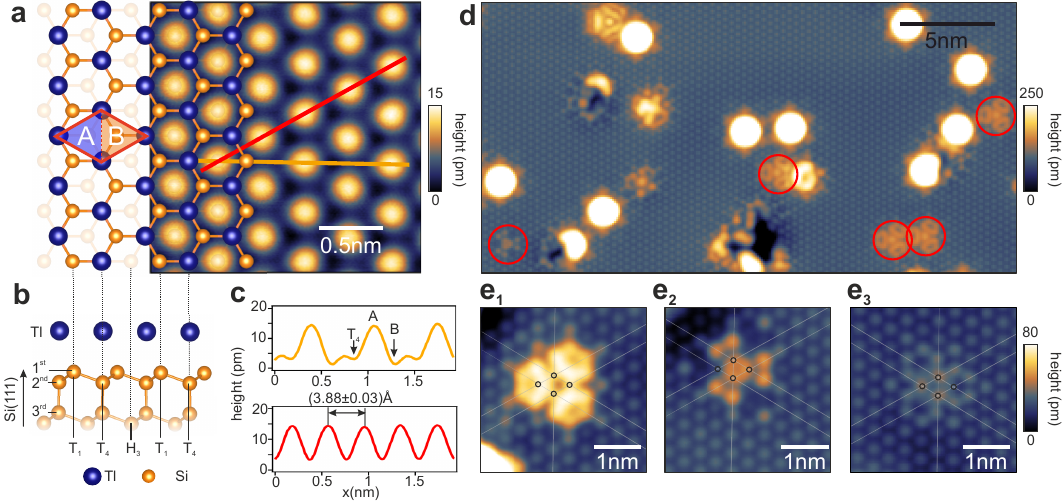}
\caption{\textbf{Topography and positional calibration of Tl/Si(111)} \textbf{a} High-resolution STM topography of Tl/Si(111) acquired at ($U_b = 650\,\text{mV}, I_T=100\,\text{pA}$), with the structural model overlaid according defect-based lattice calibration step (i). Notably, apparent height maxima in STM do not indicate Tl-atom positions, but LDOS accumulation at void position A.
\textbf{b} Side-view structural model of Tl/Si(111), showing how layer-dependent silicon substitution indicates adsorption sites T$_1$, T$_4$, H$_3$.
\textbf{c} Line profiles extracted from \textbf{a}: The orange profile reveals in-plane ISB via the inequivalent apparent height at positions A, B, and T$_4$, while the red profile confirms the $(1\times1)$ lattice constant of Tl/Si(111).
\textbf{d} Representative surface scan ($U_b = 500\,\text{mV}, I_T=30\,\text{pA}$) showing atom-sized defects with threefold rotational symmetry (encircled in red), as well as sixfold symmetric defects, likely associated with excess Tl.
\textbf{e} Zoomed-in STM images of representative threefold symmetric defects attributed to silicon-substituting phosphorous-dopant atoms ($U_b = 250\,\text{mV}, I_T=50\,\text{pA}$). Black and gray guides (hairlines and circles) mark the defect center with respect to the continued STM apparent height maxima (circles), revealing that \textbf{e}$_1$ and \textbf{e}$_2$ defects are centered on the left side of apparent height maxima, whereas the defect type shown in \textbf{e}$_3$ is centered on the right.
}
\label{Fig2}
\end{figure*}
Recently, a real-space approach based on scanning tunneling microscopy (STM) has emerged as an alternative route to access band topology, enabling the identification of atomically obstructed bands of WSe$_2$ and 1H-NbSe$_2$ \cite{Holbrook2026,Calugaru2026} and the experimental demonstration of the QSH phase in the In/SiC(0001) monolayer system indenene \cite{Bauernfeind2021,Eck2022,Erhardt2022,Schmitt2024,Erhardt2025}.
Rather than probing the band dispersion in reciprocal space, this method exploits energy-dependent charge localization arising from interference between Bloch and orbital phases contributed from neighboring lattice sites $\mathbf{R}$  

\begin{equation}
    \Psi (\textbf{r},\textbf{k}) \propto \sum_{\textbf{R}} \text{exp}(\underbrace{i\textbf{kR}}_{\text{Bloch}} \, + \, \underbrace{im\phi(\textbf{r} - \textbf{R})}_{\text{orbital}}).
    \label{equ1}
\end{equation}
Here, $m$ denotes the magnetic quantum number and $\phi(\textbf{r})$ is the angular dependence of the associated spherical harmonics, represented by the red-green-blue color sequence in Fig.~\ref{Fig1} \cite{Eck2022}.

This interference mechanism becomes particularly striking when $L_z$ eigenstates reside at the valley momenta K/K' (Fig.~\ref{Fig1}\textbf{a}). Such a situation naturally arises from in-plane inversion symmetry breaking (ISB) in a wide class of triangular lattices with $p$- or $d$-orbital character, including transition-metal dichalcogenides (TMDs), the QSH insulator In/SiC(0001) or the prototypical Rashba material Tl/Si(111) \cite{Holbrook2026,Bauernfeind2021,Sakamoto2009,Stolwijk2013,Stolwijk2014,Stolwijk2015,Stolwijk2017}.

The phase interference is illustrated in Fig.~\ref{Fig1}\textbf{b}$_1$ for a representative $m=+1$ ($p_+ = p_x + i p_y$) $L_z$ eigenstate at the K point. Without loss of generality, we consider the ISB potential at position B, thus rendering the Wyckoff positions A and B of the triangular unit cell inequivalent. Evaluating the K-point Bloch phase (gray arrows) at neighboring sites \textbf{R} introduces an \textbf{R}-dependent offset to the orbital phase $\phi(\textbf{r})$. For the $p_+$ state at K, the resulting interference is constructive at position A and destructive at site B, accumulating the local density-of-states (LDOS) at A.
Conversely, an $m=-1$ ($p_- = p_x - i p_y$) state leads to constructive (destructive) interference at B (A), as illustrated in Fig.~\ref{Fig1}\textbf{b}$_2$, with the mathematical details reported in Ref.~\onlinecite{Eck2022} and Ref.~\onlinecite{Suppl}.

Importantly, this mechanism is not impeded by the opposite $L_z$ sign of isoenergetic states at K and K' (Fig.~\ref{Fig1}\textbf{a}), imposed by time reversal symmetry conservation within the material. Since K$'=-$K, the Bloch phase likewise changes sign, yielding only a global sign change in the exponent of Eq.~(1), which leaves the real-space localization unchanged. This establishes a direct correspondence between the observed LDOS maxima and the underlying $L_z$ character.
\begin{figure*}[t!]
\includegraphics[width=18cm, height=25cm, keepaspectratio]{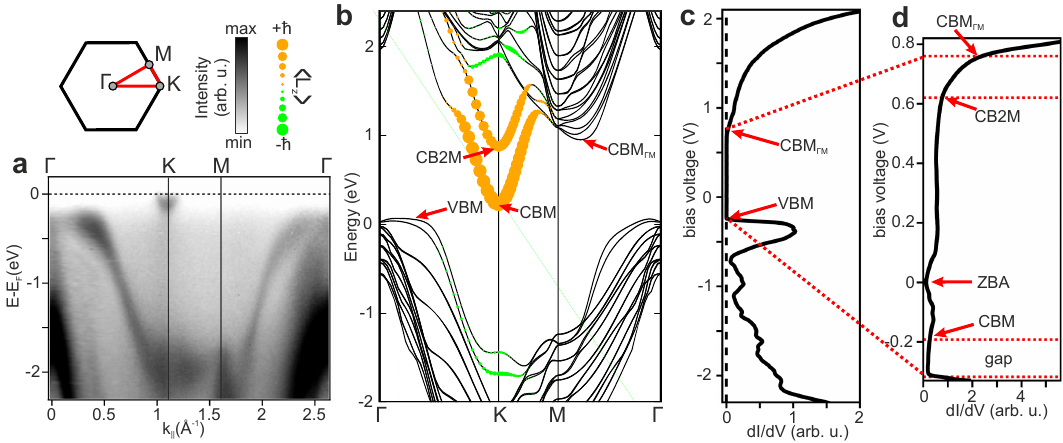}
\caption{\textbf{Electronic properties of Tl/Si(111)} 
\textbf{a} Room-temperature ARPES measurements of Tl/Si(111) acquired with photon energy $h\nu = 21.2\,\mathrm{eV}$.
\textbf{b} Band structure calculations of Tl/Si(111), with $L_z$ expectation value encoded by color. Red arrows mark relevant band onsets. 
\textbf{c} STS spectrum measured at high tunneling resistance ($U_b = -2.5\,\text{V}, I_T=1\,\text{nA}$), showing a comparatively weak tunneling contribution from in-plane K-point states relative to valence and conduction band states near the $\Gamma$ point.
\textbf{d} STS spectrum at low tunneling resistance ($U_b = -600\,\text{mV}, I_T=100\,\text{pA}$, $\Delta z = 0.1 \,\text{nm}$), where in-plane K/K'-point states are resolved; \textcolor{black}{red arrows and dashed lines denote the corresponding onsets determined from fits (see Supplementary Information \cite{Suppl}), enabling identification of the band gap and a pronounced ZBA.}
}
\label{Fig3}
\end{figure*}

In this work, we elucidate the $L_z$ sensitivity of this real-space approach from an experimental perspective by comparing two representative systems that exhibit contrasting $L_z$ band sequences. To this end, we employ a systematic workflow comprising two key steps essential for identifying $L_z$-induced off-atom charge localization: (i) precise lattice-site calibration, an indispensable step because atomic positions do not necessarily coincide with the highest-intensity features in STM topographs, rendering the exact lattice positions generally unknown. To establish the lattice registry, we use substitutional substrate dopant defects as atomic markers, supported by orbital-resolved band structure calculations \cite{Bauernfeind2021,Erhardt2025,Holbrook2026}. Then, in step (ii), the band onsets are calibrated in scanning tunneling spectroscopy (STS), thereby enabling the direct identification of the electronic bands associated with a given bias voltage. 

This workflow highlights both the strengths and limitations of the real-space approach. It provides access to occupied and unoccupied states, offering a clear advantage over dichroism in ARPES, while, at the same time, unambiguous band assignment requires either energetic non-degeneracy with other states or a dominant tunneling contribution from the target state, imposing practical constraints on the applicability of this $L_z$ detection method.

To demonstrate the $L_z$ mapping of unoccupied states, we select monolayer Tl on Si(111) as a test material (Fig.~\ref{Fig2}\textbf{a},\textbf{b}), whose unoccupied conduction-band states have remained largely unexplored with respect to their $L_z$ polarization because they are inaccessible to CD-ARPES.
In this structure, thallium atoms occupy the T$_4$ positions of the Si(111) surface, forming a well-ordered $(1\times1)$ reconstruction, as illustrated in Fig.~\ref{Fig2}\textbf{a}. STM reflects the triangular lattice symmetry of the Tl-adlayer with a lattice constant of $a=(3.88\pm0.03)$\AA$\,$ (Fig.~\ref{Fig2}\textbf{a},\textbf{c}) matching the Si(111) surface unit cell.
The structural model (Fig.~\ref{Fig2}\textbf{a},\textbf{b}) shows that the topmost Si atom reduces the symmetry to C$_{3v}$ \cite{Sakamoto2013}, thereby enabling $L_z$ polarization within the Tl $p$-orbital manifold via in-plane ISB. This symmetry reduction is directly reflected in the STM data in Fig.~\ref{Fig2}\textbf{a},\textbf{c}, where the orange line profile across the unit cell resolves three inequivalent positions, labeled A, B, and T$_4$.

For lattice site calibration, \textit{i.e.}, step (i) of our workflow, we exploit atom-sized defects observed on larger length scales (Fig.~\ref{Fig2}\textbf{d}) that only weakly perturb the Tl lattice. A subset of these lattice defects exhibits pronounced threefold symmetry and uniform orientation across the sample (Fig.~\ref{Fig2}\textbf{e}), identifying them as substitutional phosphorus dopants in subsurface silicon layers \cite{Pearson1949}. In contrast, brighter defects with apparent sixfold symmetry are attributed to excess Tl atoms, which are commonly introduced to electron-dope the Tl/Si(111) system \cite{Sakamoto2013}.
Closer inspection (Fig.~\ref{Fig2}\textbf{e}) reveals that these dopant-related defects are not centered at the apparent height maxima of the STM topography. Considering possible substitutional sites within the top three Si(111) layers, phosphorus incorporation in the first layer marks the T$_1$ site, whereas substitution in the second and third layers projects onto the T$_4$ site, as indicated in Fig.~\ref{Fig2}\textbf{b}. 

Combining these geometric considerations with the STM observations leads to the conclusion that the STM apparent height maxima correspond to the H$_3$ adsorption site of Si(111), hereafter denoted as position A. 
This assignment is supported by calculations of the charge-density localization (see Supplementary Information \cite{Suppl}) and provides an excellent example of STM probing the spatial distribution of the LDOS, whose maxima do not necessarily coincide with the atomic sites.
Finally, the defect configurations labeled \textbf{e}$_1$ and \textbf{e}$_2$ consistently mark the T$_4$ positions, thus fully calibrating the lattice positions and justifying the structural model overlay in Fig.~\ref{Fig2}\textbf{a}.

Having established the lattice positions within the triangular unit cell, we turn to step (ii), which involves a detailed examination of the Tl/Si(111) band structure, focusing on its $L_z$ polarization and the corresponding band-onset signatures in STS.
The electronic structure of Tl/Si(111) is dominated by in-plane Tl $p_{\pm}$ orbitals centered at the T$_4$ site and Si $sp_z$ orbitals allocated at T$_1$ \cite{Suppl}. This results in insulating bands with an indirect gap between the valence band maximum (VBM) at the $\Gamma$ point and the conduction band minimum (CBM) at the K/K' point, as observed by ARPES on n-doped samples with excess Tl atoms (Fig.~\ref{Fig3}\textbf{a}) \cite{Lee2002,Sakamoto2009,Sakamoto2013,Sakamoto2015}. 
Band structure calculations (Fig.~\ref{Fig3}\textbf{b}) reproduce this dispersion and reveal a strong $L_z$ character of the K point valence and conduction bands, emphasizing the similarity to Fig.~\ref{Fig1}\textbf{a}. These K-point states further exhibit pronounced Rashba splitting arising from the interplay of Tl SOC and the dominating ISB term induced by the substrate \cite{Sakamoto2009,Stolwijk2013,Stolwijk2014,Stolwijk2015,Stolwijk2017}. In the conduction bands, this leads to equal $L_z$ polarization, yet opposite spin polarization, the latter having been thoroughly demonstrated by inverse photoemission \cite{Stolwijk2013,Stolwijk2014,Stolwijk2015,Stolwijk2017}. 
Owing to its high energy position, the second conduction band (CB2) remains inaccessible to CD-ARPES, thereby requiring an alternative $L_z$-sensitive approach, such as STM-based phase-interference measurements, to probe its $L_z$ character.

Local spectroscopy (Fig.~\ref{Fig3}\textbf{c}), acquired at a large tip–sample distance, shows pronounced band onsets matching the $\Gamma$-centered VBM as well as the conduction band minimum in $\Gamma$M direction, labeled CBM$_{\Gamma \text{M}}$ (Fig.~\ref{Fig3}\textbf{b}).
These onsets delimit an energy window of vanishing differential conductance consistent with the absence of states near the $\Gamma$ point in the band calculations and reflecting tunneling suppression from the K/K' point states.
This suppression is expected from the $k$-dependence of STS \cite{Feenstra1987k}, which inherently favors tunneling from $\Gamma$-centered valence bands while damping contributions from isoenergetic K/K'-point states. This effect masks $L_z$-induced charge localization in the K/K'-point valence bands at -1.5$\,$eV and -1.7$\,$eV, restricting our analysis to the energetically isolated K/K'-point conduction bands (Fig.~\ref{Fig3}\textbf{b}).

To access these states via STS, we reduce the tunneling resistance in Fig.~\ref{Fig3}\textbf{d}, revealing finite conductance within the previously empty energy window and thereby resolving the VBM at $-320\,\text{meV}$ as well as the minimum of the first and second conduction band CBM and CB2M, respectively. 
\textcolor{black}{A detailed experimental analysis of the band onset energies based on STS, ARPES, inverse photoemission spectroscopy and QPI is presented in Supplementary Information \cite{Suppl}.}
The nearly constant LDOS up to 0.6$\,$V indicates parabolic dispersion of the first conduction band, in agreement with our band calculations in Fig.~\ref{Fig3}\textbf{b} and experimental inverse photoemission spectroscopy \cite{Stolwijk2013}. It is interrupted only by a zero-bias anomaly (ZBA), commonly observed in related thin-film materials and attributed to inelastic tunneling processes \cite{Brun2012,Odobescu2015}.
The residual conductance within the nominal gap arises from partial occupation of the K-point conduction band, as also evident in ARPES (Fig.~\ref{Fig3}\textbf{a}), and reflects a known STS artifact in degenerately doped semiconductors \cite{Feenstra1987,Bauernfeind2021,Suppl}.

\begin{figure}[t!]
\includegraphics[width=\columnwidth, height=25cm, keepaspectratio]{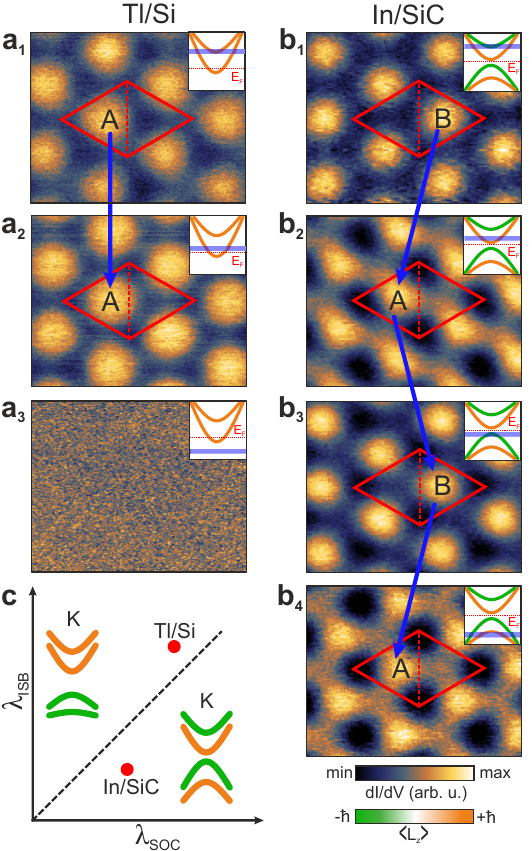}
\caption{\textbf{Band-resolved charge localization of Tl/Si(111) and In/SiC(0001)} \textbf{a}$_1$ Second and \textbf{a}$_2$ first conduction band-resolved constant-height $dI/dV$ maps of Tl/Si(111), confirming charge localization between the atoms at position A of the unit cell, as expected for equal $L_z$ character of the conduction bands at the K point. The probed band energy is indicated in the insets by a blue line. \textbf{a}$_3$ shows vanishing contrast within the gap for otherwise unchanged tunneling conditions.
\textcolor{black}{\textbf{b}$_1$--\textbf{b}$_4$ Similar constant-height $dI/dV$ measurements for the massive Dirac states of In/SiC(0001), as indicated in the inset, demonstrating alternating and off-atom LDOS maxima, corresponding to a K-point $L_z$ sequence as shown in \textbf{c}. Constant height bias voltages of \textbf{a}$_1$: $U_b = 650\,\text{mV}$, \textbf{a}$_2$: $U_b = 200\,\text{mV}$, \textbf{a}$_3$: $U_b = -250\,\text{mV}$, \textbf{b}$_1$: $U_b = 300\,\text{mV}$, \textbf{b}$_2$: $U_b = 100\,\text{mV}$, \textbf{b}$_3$: $U_b = -100\,\text{mV}$, \textbf{b}$_4$: $U_b = -440\,\text{mV}$ were applied.}
\textbf{c} ISB--SOC phase diagram of the triangular $p$-orbital lattice relevant to Tl/Si(111) and In/SiC(0001), where $\lambda_{\text{ISB}}$ and $\lambda_{\text{SOC}}$ denote the ISB and SOC strengths, respectively. The indicated $L_z$ sequences and corresponding LDOS maxima show that the electronic structure of Tl/Si(111) is dominated by ISB, whereas SOC constitutes the stronger term for In/SiC(0001).
}
%
\label{Fig4}
\end{figure}

With both the lattice position and band onsets of Tl/Si(111) established, we are now in a position to investigate $L_z$-dependent charge localization within the unit cell. A constant-height differential conductance map acquired at a bias of $650\,\text{mV}$ (Fig.~\ref{Fig4}\textbf{a}$_1$) reveals the LDOS maximum at position A and, according to our STS calibration in step (ii), corresponds to the second conduction band CB2. 
Lowering the bias to the first conduction band CB ($200\,\text{mV}$, Fig.~\ref{Fig4}\textbf{a}$_2$) leaves the LDOS maximum unchanged in position, consistent with identical $L_z$ character in both conduction bands. Systematic measurements across the bands (see Supplemental Information \cite{Suppl}) confirm that the LDOS maximum does not shift with energy in the first and second conduction band window of Fig.~\ref{Fig3}\textbf{d}, while the signal vanishes within the gap (Fig.~\ref{Fig4}\textbf{a}$_3$). 
The persistence of the charge maximum at position A, located between the Tl atoms, provides a clear fingerprint of identical $L_z$ character of the K-point conduction bands, precisely as predicted by theory (Fig.~\ref{Fig3}\textbf{b}) \cite{Suppl}. This indicates that for Tl/Si(111) ISB dominates over the large atomic thallium SOC (Fig.~\ref{Fig4}\textbf{c}), which we attribute to enhanced ISB due to the T$_4$ adsorption geometry. 
Notably, the LDOS maximum resides at the hollow H$_3$ site, where neither thallium nor silicon atoms of the surface unit cell are located, highlighting the role of phase interference in the context of atomic obstruction \cite{Holbrook2026,Eck2023}.

To further substantiate the $L_z$ dependence of this localization mechanism, we perform analogous site and band calibration as well as $dI/dV$ measurements on monolayer In/SiC(0001) films (Fig.~\ref{Fig4}\textbf{b}) \cite{Suppl}. The  K/K'-point band structure of this QSH insulator dubbed indenene likewise arises from $p_+$ and $p_-$ orbitals subject to ISB and SOC \cite{Bauernfeind2021,Eck2022,Erhardt2025}. 
In contrast to Tl/Si(111), the corresponding LDOS maxima in Fig.~\ref{Fig4}\textbf{b} alternate between position B and A across different Dirac bands and can even be traced into the valence band, facilitated by the absence of energetically competing indenene states at lower crystal momenta $\textbf{k}$. Denser bias voltage sampling of the In/SiC(0001) K/K'-point states \textcolor{black}{as well as data of higher n-type indenene, with partially occupied conduction band, similar to Tl/Si(111) are} presented in the Supplementary Information \cite{Suppl}. This behavior is consistent with the alternating $L_z$ character of the ISB-SOC phase diagram in Fig.~\ref{Fig4}\textbf{c} and arises from a SOC-driven topological band inversion in In/SiC(0001), promoted by a reduced ISB term \cite{Bauernfeind2021,Eck2022,Erhardt2025}.


In conclusion, the comparison between Tl/Si(111) and In/SiC(0001) establishes the $L_z$ dependence of real-space LDOS localization. It demonstrates that STM can access the orbital polarization of occupied \textit{and} unoccupied K/K'-point states, provided that the atomic lattice sites and energy positions are accurately calibrated. Our study further allows to identify Tl/Si(111) as the topologically trivial counterpart of the QSH insulator In/SiC(0001) within the phase diagram of triangular $p$-orbital systems.


The applicability range of this approach is broad. As mentioned in the introduction, the obstructed-atomic limit in TMDs has been demonstrated recently \cite{Holbrook2026,Calugaru2026}. There, the $d$-wave nature of the planar orbitals gives rise to the \textit{same} interference effects described here for $p_\pm$, simply with higher orbital angular momentum.
Another class of systems where this physics can be probed is multilayer graphene, especially twisted bilayer graphene (TBG). The key difference to our case in point are the inter-moiré distances that are about ten times larger than in the present non-moiré systems. 
STM has been widely applied to TBG \cite{Nuckolls2020,Nuckolls2023,Wong2020,Xie2019,Kim2023,Kim2026,Tong2022} with partial focus on the real-space localization of the moiré orbitals. Extending our orbital-based analysis to TBG could refine the comparison between the various localized orbital bases proposed within the different lattice versions \cite{Song2022,Koshino2018,Haule2019,Carr2019} 
of the Bistritzer-MacDonald continuum model \cite{Bistritzer2011}, thereby 
facilitating a more transparent lattice representation of the rich 
TBG physics.